\documentstyle[prl,aps,twocolumn]{revtex}

\begin{document}

\title{Can a local repulsive potential trap an electron?}
\author{N.\ Berglund}
\address{Institute of Theoretical Physics, Ecole Polytechnique
F\'ed\'erale de Lausanne, PHB-Ecublens, CH--1015 Lausanne,
Switzerland}
\author{Alex Hansen, E.\ H.\ Hauge}
\address{Department of Physics,
Norwegian University of Science and Technology,
         N--7034 Trondheim, Norway}
\author{J.\ Piasecki}
\address{Institute of Theoretical Physics, University of Warsaw, Ho\.{z}a 69,
          PL--00\, 681 Warszawa, Poland}
\date{\today}
\maketitle

\begin{abstract}

We study the classical dynamics of a charged particle in two dimensions,
under the
influence of a perpendicular magnetic and an in-plane electric field.
We prove the surprising fact that there is a finite region
in phase space that corresponds to the otherwise drifting
particle being trapped by a local {\it repulsive} potential.
Our result is a direct consequence of KAM-theory and, in
particular, of  Moser's theorem. We illustrate it by numerical
phase portraits and by an analytic approximation to invariant
curves.

\end{abstract}


The detailed dynamics of electrons in two dimensions (2D), in
electromagnetic fields and with localized scatterers, is important
for several aspects of the quantum Hall effect. Classical dynamics
is more relevant in this context than one might think, see the recent review
by Trugman\cite{trugman}. In addition, classical kinetic theory
for such systems is surprisingly subtle\cite{gbe}, and requires for
its foundation an understanding of the underlying dynamics.
In this letter we focus on the {\it classical} dynamics
of an ``electron" (i.e., a particle with charge $-e$) confined to the
$xy-$plane. Perpendicular to these two dimensions
there is a constant magnetic field $\vec{\cal B}=
{\cal B}{\bf e}_z$. In addition, there is a constant in-plane electric field
$\vec{\cal E} ={\cal E}{\bf e}_y$.
When no forces act on the electron except those stemming from these
fields, the nature of the motion is well known. With ${\cal E}=0,\,
{\cal B}\neq 0$, the electron with energy $E={\textstyle \frac{1}{2}}
mv^2$ moves in a circular orbit with cyclotron radius $R=v/\omega_c$
and cyclotron frequency $\omega_c =e{\cal B}/m$. Addition of a finite
electric field results in the center of this orbit
acquiring  a constant drift velocity
$\vec{v}_d=(\vec{\cal E}\times \vec{\cal B})
/{\cal B}^2=({\cal E}/{\cal B}){\bf e}_x$.
The question we address in this letter is the following: With both
$\vec{\cal E}$- and $\vec{\cal B}$-field present, is there a finite
region in phase space which corresponds to the otherwise
drifting electron being trapped by  a single scatterer, i.e., by a local
{\it repulsive} potential?
(For concreteness, we let the repulsive potential be a hard disk
of radius $a$.)
This question arises, for example, when one tries to construct
a kinetic theory for the Lorentz gas in 2D, with both electromagnetic
fields present.

When $\vec{\cal E}=0,\, \vec{\cal B}\neq 0$
all electrons are, in a sense, trapped by the
magnetic field, and the hard disk simply modifies the cyclotron orbits
into orbits skipping around the periphery of the disk
(see for instance \cite{bk}). On the other
hand, when $\vec{\cal E}\neq 0,\, \vec{\cal B}=0$, one or more collisions with
the disk will only temporarily perturb the
dynamics determined by a constant acceleration in the negative
electric field direction -${\bf e}_y$. With both fields present,
the question is whether an electron initially colliding with the
disk, will ultimately miss it, due to the drift caused by the
electric field.

There are two dimensionless parameters in the
problem, the ratio of the cyclotron radius and that of the
disk, $r\equiv R/a$, and the ratio of the displacement, $v_d\cdot
 2\pi /\omega_c$, during one cyclotron
period and the radius of the disk,
$\varepsilon \equiv (2\pi m/ae)\cdot ({\cal E}/{\cal B}^2)$.
Clearly, the answer to our question whether the disk can trap
the electron when drift is included, must depend on where
in this two-dimensional parameter space it is posed.
Our aim in this letter is not to give a complete answer to the question,
with parameter regions and corresponding measures precisely
delineated. Our more modest aim is to demonstrate the surprising
{\it qualitative} fact that a repulsive potential
can, indeed, trap a charged particle in crossed electromagnetic fields
in 2D.

We first emphasize that an electron drifting towards the hard
disk from ``infinity" cannot be trapped. The main elements
of the proof are:
To the first collision with the disk is associated a finite measure
in phase space. Consider the subset of those incoming
trajectories (if they exist) that will hit the disk at least
$N$ times. Since classical dynamics is measure preserving,
the regions in phase space associated with each of the $N$ collisions
have the same measure. Reversibility (invariance under reversal of time
and magnetic field) implies that they do not overlap.
Since the total phase space associated with an electron colliding with the disk
is finite, it follows that the measure of the trajectories coming from infinity
and hitting the scatterer at least $N$ times has to converge to zero
as $N\to \infty$. (It may, of course, vanish for a finite $N$.)
Trapping is, therefore,
only possible for electrons that are initially close to the scatterer.

Naively one might think that conditions are more favorable for
trapping when $r\ll 1$. If this were so, not much room
would be left for trapping: When $r\ll 1$, the perimeter of the disk can, on
the
scale set by the cyclotron radius, be considered a straight line.
An electron skipping upwards along the right side of the
disk will, unless the curvature of the disk makes itself felt
in time, sooner or later miss the disk and drift away from it.

Surprisingly, it is the opposite regime, when $r\gg 1$, which is
more favorable for trapping. Here we can, on the scale of the
radius of the disk, consider the parts of the electron orbit
close to the disk as straight lines. Two subsequent collisions are
shown in Fig.\ref{f1}. The direction of the motion just prior to
collision no. $n$ is defined by the angle  $\phi_n$. To the extent that the
orbit
locally can be considered to consist of straight lines, the scattering angle of
collision no. $n$ is given by $\psi_n=\phi_{n+1}-\phi_n$.
Due to the electric field, the incoming line heading for
collision no. $(n+1)$ is shifted by the amount $\varepsilon a$
in the positive $x$-direction with respect to the outgoing
line from collision no. $n$. In terms of the angles $\phi_n$ and the
dimensionless impact parameters $\beta_n=b_n/a$, inspection
of Fig.\ref{f1} gives the map,
\begin{eqnarray}  \label{1}
  \phi_{n+1}&=&\phi_n +\pi -2\sin^{-1}\beta_n \nonumber \\
  \beta_{n+1}&=&\beta_n-\varepsilon \sin \phi_{n+1}.
\end{eqnarray}
In contrast to standard scattering conventions in three dimensions,
inherent in the present map is the following choice:
The impact parameter $\beta$ lives on the interval $(-1,1)$
and the scattering angle $\psi$ on the interval
$[0, 2\pi)$, with $\beta \to -1$ corresponding to $\psi\to 2\pi$.

One can show that the map resulting from the full dynamics for arbitrary
$\cal B$ reduces to the map (\ref{1}) when the time interval between
successive collisions is approximated by the cyclotron period $2\pi /\omega_c$.
This implies that corrections to (\ref{1}) are uniformly bounded by
a constant times $a/R=r^{-1}$. Consequently, the map (\ref{1}) is reliable when
$r\gg 1$. 

The map (\ref{1}) has several interesting properties. First of all,
since its Jacobian equals unity, it is area-preserving.
Thus, our weak $\cal B$-field  approximation, basic to (\ref{1}),
has exactly preserved this general property of Hamiltonian maps.
Next, for small $\varepsilon$,  (\ref{1}) is a small perturbation
of the integrable map
\begin{eqnarray}  \label{2}
  \phi_{n+1}&=&\phi_n + \Omega(\beta_n) \nonumber \\
  \beta_{n+1}&=&\beta_n,
\end{eqnarray}
where $\Omega(\beta) = \pi -2\sin^{-1}\beta$ describes scattering off a
hard disk.
{From} (\ref{2}) we recover the well-known fact that the scattering angle
$\psi_n = \phi_{n+1} - \phi_n$ remains the same in subsequent collisions,
when the electric field vanishes \cite{bk}.
Furthermore, $\Omega(\beta)$ is a
monotonic function of its argument and, as a result,
our map (\ref{1}) belongs to the
category of twist maps \cite{meiss}, for which a number of
properties have been rigorously established.

However, the question of principal interest from our point of
view is the existence of trapped orbits, and the
phase space associated with them, when $\varepsilon>0$.
The function $\Omega(\beta)$ is undefined when
$|\beta|>1$. Physically,
this corresponds to the electron missing the disk.
The regions of first and last collision are represented
in phase space by areas of order $\varepsilon$ near
$\beta = +1$ and $\beta = -1$. The nontrivial
question is, therefore, whether there is a finite
region of phase space associated with orbits that
{\it never}
enter the escape region.

For small enough $\varepsilon$, Kolmogorov--Arnold--Moser (KAM) theory enables us to
give a positive answer to this question. Specifically, in our case of
a two-dimensional map, we will use Moser's
theorem\cite{moser} which proves the existence of invariant curves in
phase space for small perturbations, and for some initial
conditions. The existence of two distinct invariant curves
winding around phase space,
together with area preservation, shows the 
finite region enclosed between these two curves to be an invariant
set, corresponding to trapped orbits.
It is straightforward to verify that the map (\ref{1}) (or any
small perturbation of it that remains area preserving)
satisfies the hypotheses of
Moser's theorem, provided that we restrict phase
space to a smaller strip $|\beta|<c<1$, in order to
avoid the escape regions and the singularities of
$\Omega$. The important point is that this theorem  does
{\em not} require the entire strip to be invariant under the map.
We have thus proved the existence, for small electric and magnetic fields,
of invariant curves in phase space, and hence of trapped orbits 
with positive measure.

This rigorous result is illustrated numerically
by figures \ref{f2} and \ref{f3}, where
we show two different phase portraits of the map.
When $\varepsilon = 0.2$, many invariant curves are present.
We can divide them into two classes: those winding around
phase space, which we study in this letter, and those
forming ``elliptic islands" around periodic orbits.
When $\varepsilon = 0.4$, the structure
of phase space has become
more complex. Invariant curves winding around phase space
are now scarce and, numerically, all of them
seem to be destroyed
between $\varepsilon = 0.4$ and $\varepsilon = 0.5$. 
On the other hand, elliptic islands, which are numerous in
Fig.\ref{f3}, survive at much stronger electric fields
(certainly up to $\varepsilon\sim 1.1$). However, most of the 
chaotic trajectories, which remained bounded for
small $\varepsilon$, diffuse until they reach the escape region.

As an analytic illustration of our rigorous result, we
have also constructed a perturbation scheme around (\ref{2}), close in spirit
to that used in the proof of Moser's theorem\cite{moser}.
We start from the Ansatz\cite{motive},
\begin{eqnarray}  \label{3}
  \phi_n&=&\zeta +n\omega +\varepsilon u_1(n;\zeta ,\omega)+\varepsilon^2
   u_2(n;\zeta ,\omega)+\cdots  \\
  \psi_n&=&\phi_{n+1}-\phi_n=\omega +\varepsilon v_1(n;\zeta , \omega)
   +\varepsilon^2v_2(n;\zeta ,\omega)+\cdots , \nonumber
\end{eqnarray}
with $v_i(n)=u_i(n+1)-u_i(n)$. Both the ``effective initial condition" $\zeta$, and
the ``angular frequency" $\omega$ depend on $\varepsilon$
and on the true initial conditions $(\phi_0,\psi_0)$. Setting $n=0$ in
(\ref{3}) gives the inverse relations $\phi_0 =\phi_0(\zeta ,\omega ;\varepsilon),\
\psi_0=
\psi_0(\zeta , \omega ;\varepsilon)$ via the functions $u_i(0;\zeta ,\omega),\
v_i(0;\zeta ,\omega)$.
We determine $u_i(n)$ and $v_i(n)$ by inserting the Ansatz (\ref{3}) into
the map (\ref{1}) and doing straightforward perturbation theory, treating
$\zeta$ and $\omega$ as constants in the process. We fix the ``initial values"
$u_i(0)$ and $v_i(0)$ by {\it requiring} that the oscillating functions
$u_i(n)$ and $v_i(n)$ average to zero for large $n$. (This is indeed
necessary for the Ansatz to be consistent.)
Convergence of the scheme is assured\cite{moser} if $\omega/2\pi$ is a
Diophantine number (loosely speaking: an irrational number which
is hard to approximate by a sequence of rationals).

To zeroth order the
result is simple, $\zeta =\phi_0,\ \omega =\psi_0$, reproducing the 
skipping orbits in zero electric field. To first order, perturbation
theory gives
\begin{eqnarray} \label{4}
  v_1(n)&=&\frac{1}{\sin^2(\omega/2)}\left\{ \cos[\zeta +\omega/2] \right. 
      \nonumber \\     & & \left. -\cos[\zeta +
   (n+{\textstyle \frac{1}{2}})\omega ]\right\} +v_1(0).
\end{eqnarray}
In (\ref{4}) we determine $v_1(0)$ by requiring $v_1(n)$ to average
to zero for large $n$. We can then calculate $u_1(n)$ by direct summation,
again fixing $u_1(0)$ by the stipulation that $u_1(n)$ averages to zero.
The results are,
\begin{equation}   \label{5}
  v_1(0)=-\frac{\cos (\zeta +\omega/2)}{\sin^2(\omega/2)}\ \ ; \ \
  u_1(0)=-\frac{\sin \zeta}{2\sin^3(\omega/2)}.
\end{equation}
In this way one can continue. We are primarily
interested in the frequency $\omega(\varepsilon)$.
To second order, we determine $v_2(0)$ in the same manner, and contruct the
inverse relation from (\ref{3}) to ${\cal O}(\varepsilon^2)$  as
\begin{equation} \label{6}
  \psi_0=\omega -\varepsilon\frac{\cos (\zeta+\omega/2)}{\sin^2(\omega/2)}+
    \varepsilon^2\frac{\cos (2\zeta+\omega)}{8\sin^3(\omega/2)\cos (\omega/2)}.
\end{equation}
Inversion of this relation (with appeal to the first order result
for $\zeta$) to ${\cal O}(\varepsilon^2)$ finally yields
\begin{eqnarray} \label{7}
\omega & = & \psi_0 + \varepsilon  \frac{\cos(\phi_0+\psi_0/2)}
{\sin^2(\psi_0/2)} 
 -  \varepsilon^2 \left[ \frac{\cos(2\phi_0+\psi_0)}
{8\sin^3(\psi_0/2)\cos(\psi_0/2)} \right. \nonumber \\ & &\ \left. +
\frac{\cos(\psi_0/2)[\cos(2\phi_0+\psi_0)+3]}
{4\sin^5(\psi_0/2)} \right] + {\cal O}(\varepsilon^3).
\end{eqnarray}
Note that in the second order term of (\ref{6}), there appears a ``small
denominator", $\cos(\omega/2)$, which vanishes at $\omega=\pi$, i.e.,
$\beta=0$ (where an orbit of period two exists). Higher
order terms will generate a proliferation of other small denominators.
This denominator appearing in second order is the first symptom
indicating that convergence of the scheme requires $\omega/2\pi$ to be
a Diophantine number. To see that the solution indeed
describes an invariant curve, it is sufficient to note that, by construction,
$v_i(n;\zeta ,\omega)=v_i(0;\zeta +n\omega , \omega)$, with a
similar equation for $u_i$. Thus, for given $\varepsilon$ and $\omega$,
$\phi_n$ and $\psi_n$ depend only on the combination $\zeta +n\omega$. Elimination of this
parameter and use of (\ref{2}) yields an explicit equation for the
invariant curve, $\beta_n =\Omega^{-1}(\psi_n)=f(\phi_n;\omega ,\varepsilon)$.
As stated above, once two distinct invariant
curves have been constructed, they delimit an invariant region of
positive measure corresponding to trapped electrons.

For Diophantine $\omega/2\pi$, when the perturbation scheme implied
by (\ref{3}) converges, by taking the average one concludes that $\langle \psi_n 
\rangle =\omega$.
In Fig.\ref{f4} we show a comparison of the scattering angle
averaged over 1\,000\,000 collisions, and our analytic
perturbation result for the frequency, with the 
initial conditions $(\phi_0 = 0.3, \psi_0=2.0)$.
For small $\varepsilon$ the results are in very good agreement,
despite the fact that $\omega$ takes both rational and irrational
values.  
For larger $\varepsilon$, one can observe the breakdown of
perturbation theory, due to the phenomenon of ``mode
locking": 
On the interval of roughly $0.20<\varepsilon <0.45$
the electron is trapped in the vicinity of a periodic orbit
with period 3 and, as a consequence, $\langle \psi_n \rangle /
2\pi =1/3$, a rational number.

In conclusion, we have shown that for sufficiently small
electric and magnetic fields, bound states associated with
a hard disk scatterer constitute a set of positive measure.
Indeed, both fields add small and smooth perturbations
to the integrable map (\ref{2}), so that KAM theory
applies. By investigating the stability of periodic
orbits, it is in fact possible to show that trapping occurs also
for moderate values of the magnetic field. Note, finally, that
generalization from a hard disk to an arbitrary, rotationally invariant
repulsive potential of strictly finite range,
leaves the scattering function $\Omega(\beta)$ monotonic.
In other words, the corresponding map is still a twist
map, and all qualitative conclusions remain unchanged.
In short, we have accomplished our goal and, within
classical mechanics, answered the
question:{\it Yes}, a local repulsive potential {\it can} trap an electron!
With quantum mechanics, the question remains open.

{\it Acknowledgements.} We would like to thank Professors  J. S. H\o ye and H. Kunz 
for important remarks. N.B., E.H.H. and J.P. are grateful
to Professor J.-P. Hansen and the Ecole Normale Sup\'{e}rieure de Lyon,
for their hospitality. Finally, N.B. ackowledges financial support
from the Swiss National Fund, J.P., A.H. and E.H.H. from The Norwegian
Research Council and J.P. from KBN (The Polish Committee for
Scientific Research), grant no. 2~P302 074 07.

\begin{figure}
\caption{
 Geometry of two successive collisions with the hard disk.}
\label{f1}
\end{figure}

\begin{figure}
\caption{
Phase portrait of the mapping for $\varepsilon =0.2$. The escape
regions are recognized as those close to the upper and lower
part of the figure. Note that orbits of period 2, 3 and
4 are easily located. Between the periodic orbits are
bands of invariant curves, described by the analytic theory
of the text.}
\label{f2}
\end{figure}

\begin{figure}
\caption{
Same as Fig.2, but for $\varepsilon= 0.4$. The escape
regions have grown in size, more periodic orbits can be
recognized, and the regions with invariant curves have shrunk
considerably. In addition, chaotic regions (which remain trapped!)
have become manifest.}
\label{f3}
\end{figure}

\begin{figure}
\caption{
Comparison between the ``frequency" 
$\omega/2\pi$ as computed from
our analytic approximations to first and second order,
and $\langle \psi_n\rangle/2\pi$ obtained from 
the simulation, where the average has
been taken over 1\,000\,000 collisions, starting from
the initial condition
$\phi_0=0.3, \psi_0=2.0$.
Note, with reference to Fig.2, that the electron is mode locked to
a period 3 orbit, $\langle \psi_n\rangle/2\pi =
1/3$, from $\varepsilon\approx 0.2$ to $\varepsilon\approx 0.45$.}
\label{f4}
\end{figure}


\begin{references}
\bibitem{motive} Some readers may find the following 
motivation for the Ansatz useful:
An invariant curve has a parametric equation of the form
$\phi = \xi + \varepsilon U(\xi,\omega,\varepsilon)$, $\psi = 
\omega + \varepsilon V(\xi,\omega,\varepsilon)$.
The parameter $\xi$ grows proportionally to time with ``angular
frequency'' $\omega$: $\xi_n = \zeta + n \omega$.
\end{references}
\end{document}